\documentclass[showpacs,aps,graphicx,twocolumn]{revtex4}%
\usepackage{graphicx}
\begin{document}
\title{Symmetric multiparty-controlled teleportation of an arbitrary two-particle
entanglement}
\author{ Fu-Guo Deng,$^{1,2,3}$\footnote{
 E-mail addresses: fgdeng@bnu.edu.cn} Chun-Yan Li,$^{1,2}$
 Yan-Song Li,$^{4}$\footnote{
 E-mail addresses:  ysli@tsinghua.edu.cn}
 Hong-Yu Zhou,$^{1,2,3}$\footnote{
 E-mail addresses: zhy@bnu.edu.cn} and Yan Wang$^{1,2}$ }
\address{$^1$ The Key Laboratory of Beam Technology and Material
Modification of Ministry of Education, Beijing Normal University,
Beijing 100875, People's Republic of China\\
$^2$ Institute of Low Energy Nuclear Physics, and Department of
Material Science and Engineering, Beijing Normal University,
Beijing 100875, People's Republic of China\\
$^3$ Beijing Radiation Center, Beijing 100875,  People's Republic
of China \\
$^4$ Department of Physics, and Key Laboratory For Quantum
Information and Measurements of Ministry of Education, Tsinghua
University, Beijing 100084, People's Republic of China}
\date{\today }

\begin{abstract}
We present a way for symmetric multiparty-controlled teleportation
of an arbitrary two-particle entangled state based on Bell-basis
measurements by using two Greenberger-Horne-Zeilinger states,
i.e., a sender transmits an arbitrary two-particle entangled state
to a distant receiver, an arbitrary one of the $n+1$ agents via
the control of the others in a network. It will be shown that the
outcomes in the cases that $n$ is odd or it is even are different
in principle as the receiver has to perform a controlled-not
operation on his particles for reconstructing the original
arbitrary entangled state in addition to some local unitary
operations in the former. Also we discuss the applications of this
controlled teleporation for quantum secret sharing of  classical
 and  quantum information. As all the instances can be
used to carry useful information, its efficiency for qubits
approaches the maximal value.
\end{abstract}
 \maketitle

\section{introduction}

The principles of quantum mechanics supplied many interesting
applications in the field of information in the last decade, such
as quantum computer, quantum cryptography, quantum teleportation,
quantum secret sharing, and so on. The quantum teleportation
process allows the two remote parties, the sender Alice and the
receiver Bob, to utilize the nonlocal correlations of the quantum
channel, Einstin-Podolsky-Rosen (EPR) \cite{EPR} pair shared
initially, to teleport an unknown quantum state $\vert
\chi\rangle=a\vert \uparrow\rangle + b\vert \downarrow \rangle$;
Alice makes a Bell-basis measurement on her EPR particle and the
unknown quantum system $\chi$, and Bob reconstructs the state
$\vert \chi\rangle$ with a local unitary operation on his EPR
particle according to the classical information published by Alice
\cite{teleportation}. Quantum teleportation has been demonstrated
by some groups \cite{TE1,TE2,TE3,TE4} since Bennett et al.
\cite{teleportation} proposed the theoretical protocol for
teleporting an unknown single qubit in 1993. Subsequently, the
protocols for teleporting an entangled state are proposed with
some pure entangled states or maximal multiparticle entangled
states
\cite{TT1,TT2,ShiGuo,YangGuo,LuGuo,Lee,YanFLteleportation1}. For
example, Lu and Guo \cite{LuGuo} introduced some ways for
teleporting an entangled state $\alpha \vert 00\rangle+\beta\vert
11\rangle$ with entanglement swapping
\cite{swapping1,swapping2,swapping3} by using EPR pairs or pure
entangled states as the quantum channels in 2000. Lee proposed a
protocol \cite{Lee} for teleporting an entangled state $\alpha
\vert 10\rangle+\beta\vert 01\rangle$ with the four-particle
Greenberger-Horne-Zeilinger (GHZ) state $\vert
\psi\rangle_{L}=\frac{1}{\sqrt{2}}(\vert 1010\rangle+\vert
0101\rangle)$. Recently, Rigolin \cite{Rigolin} showed a way to
teleport an arbitrary two-qubit entangled state with a
four-particle entangled state $\vert
\psi\rangle_R=\frac{1}{2}(\vert 0000\rangle+\vert
0101\rangle+\vert 1010\rangle+\vert 1111\rangle)$ and
four-particle joint measurements.

Quantum secret sharing (QSS) is an important branch of quantum
communication and is used to complete the task of classical secret
sharing with the principles of quantum mechanics. The basic idea
of secret sharing \cite{Blakley} in a simple case (there are three
parties of communication, say Alice, Bob, and Charlie) is that a
secret is divided into two pieces which will be distributed to two
parties, respectively, and they can recover the secret if and only
if both act in concert. A pioneering QSS scheme was proposed by
Hillery, Bu\v{z}ek, and Berthiaume \cite{HBB99} in 1999 by using
the three-particle and four-particle entangled GHZ states for
sharing  classical information. Now, there are a lot of works
focused on QSS in both the theoretical
\cite{HBB99,KKI,Bandyopadhyay,Karimipour,guoqss,longqss,qssPLADLZ,delay,Peng,MZ,DZL,cleve}
and experimental \cite{TZG,AMLance} aspects.  Different from
classical secret sharing, QSS can be used to sharing both
classical and quantum information. For instance, the QSS protocols
in Refs. \cite{HBB99,KKI,cleve,Bandyopadhyay,Peng} are used to
split a quantum secret.

Recently,  controlled teleporation for a single-qubit $\vert
\chi\rangle=a\vert \uparrow\rangle + b\vert \downarrow \rangle$
\cite{CTele1,YanFLteleportation2} or m-qubit message
$\prod_{i=1}^{m} \otimes(\alpha_i\vert 0\rangle_i+\beta_i\vert
1\rangle_i)$ \cite{CTele2} have been studied. In those
teleportation protocols, the qubits can be regenerated by one of
the receivers with the help of the others. Those principles can be
used to split a quantum secret in QSS \cite{HBB99}. In this paper,
we will present a symmetric protocol for multiparty-controlled
teleportation of an arbitrary two-particle entangled state with
two GHZ states and Bell-basis measurements. It can be used to
share a classical information and an entangled quantum secret.
Different from the protocols for teleportation of a two-particle
entangled state with a GHZ state in which the unknown state should
be an EPR-class entangled state \cite{TT1,TT2}, i.e., $\vert
\Phi\rangle_{\chi}=\alpha\vert uv\rangle + \beta \vert
\bar{u}\bar{v}\rangle$ ($u,v\in \{0, 1\}$, and $\bar{u}=1-u$), the
unknown quantum system in this protocol is in an arbitrary
two-particle state. Moreover, the receiver is an arbitrary one in
the $n+1$ agents via the control of the others in the network. As
the  whole quantum source is used to carry the useful quantum
information, the efficiency for the qubits approaches the maximal
value and the procedure for  controlled teleportation is an
optimal one.

The paper is organized as follows. In Sec. II, we present a way
for the symmetric controlled teleportation of an arbitrary
two-particle entangled state with two three-particle GHZ states.
That is, there is one controller who controls the process of
quantum teleportation. We generalize it to the case with $n+1$
agents in which one is the receiver and the other $n$ agents are
the controllers in the network in Sec. III, and discuss the
difference between the two cases where the number of controllers
is even or odd. In Sec. IV, we apply the method for the controlled
teleportation to share  classical  and  quantum information. A
brief discussion and summary are given in Sec. V.


\section{controlled teleportation via the control of one agent}

An EPR pair is in one of the four Bell states shown as follows
\cite{book}:
\begin{eqnarray}
\left\vert \psi ^{\pm}\right\rangle_{AB}
=\frac{1}{\sqrt{2}}(\left\vert 0\right\rangle _{A}\left\vert
1\right\rangle _{B}\pm\left\vert
1\right\rangle _{A}\left\vert 0\right\rangle _{B}), \label{EPR12}\\
\left\vert \phi ^{\pm}\right\rangle_{AB}
=\frac{1}{\sqrt{2}}(\left\vert 0\right\rangle _{A}\left\vert
0\right\rangle _{B}\pm\left\vert 1\right\rangle _{A}\left\vert
1\right\rangle _{B}), \label{EPR34}
\end{eqnarray}
where $\vert 0\rangle$ and $\vert 1\rangle$  are the eigenvectors
of the operator $\sigma_z$. The four unitary operations $\{U_i\}$
($i=0,1,2,3$) can transfer each one of the four Bell states into
another,
\begin{eqnarray}
U_{0}=\left\vert 0\right\rangle \left\langle 0\right\vert
+\left\vert 1\right\rangle \left\langle 1\right\vert, \,\,\,\,\,
U_{1}=\left\vert 0\right\rangle \left\langle 0\right\vert
-\left\vert 1\right\rangle \left\langle 1\right\vert, \nonumber\\
U_{2}=\left\vert 1\right\rangle \left\langle 0\right\vert
+\left\vert 0\right\rangle \left\langle 1\right\vert, \,\,\,\,\,\,
U_{3}=\left\vert 0\right\rangle \left\langle 1\right\vert
-\left\vert 1\right\rangle \left\langle 0\right\vert. \label{U}
\end{eqnarray}

Suppose the unknown two-particle state teleported is
\begin{eqnarray}
\vert \Phi\rangle_{xy}=a\vert 00\rangle_{xy} + b
\vert 01\rangle_{xy} + c\vert 10\rangle_{xy} + d\vert
11\rangle_{xy},\label{unknownstate}
\end{eqnarray}
where
\begin{eqnarray}
\vert a\vert^2 + \vert b\vert^2 + \vert c\vert^2 + \vert d\vert^2
=1,
\end{eqnarray}
and the three-particle GHZ state prepared by Alice is
\begin{eqnarray}
\vert GHZ\rangle_{ABC}=\frac{1}{\sqrt{2}}(\vert 000\rangle+\vert
111\rangle).
\end{eqnarray}
 With a Hadamard (H) operation on each particle,
\begin{equation}
H=\frac{1}{\sqrt{2}}\left(\begin{array}{cc}
1 & 1\\
1 & -1
\end{array}
\right),
\end{equation}
the state becomes
\begin{eqnarray}
\vert GHZ'\rangle_{ABC}=\frac{1}{\sqrt{2}}(\vert
+x+x+x\rangle+\vert -x-x-x\rangle),
\end{eqnarray}
where $\vert +x\rangle=\frac{1}{\sqrt{2}}(\vert 0\rangle+\vert
1\rangle)$ and $\vert -x\rangle=\frac{1}{\sqrt{2}}(\vert
0\rangle-\vert 1\rangle)$ are the two eigenvectors of the operator
$\sigma_x$.


\begin{figure}[!h]
\begin{center}
\includegraphics[width=8cm,angle=0]{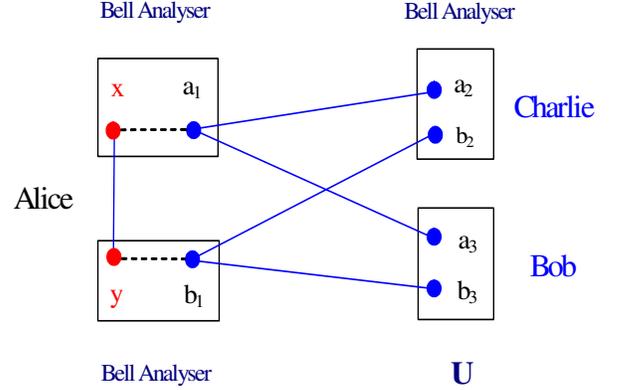} \label{fig1}
\caption{ (Color online) Symmetric controlled teleportation of an
arbitrary two-particle entangled state with two GHZ states. Alice,
Bob, and Charlie each keep one of the three particles in each GHZ
state. The bold lines connect qubits in GHZ states or the
two-particle arbitrary entangled state $\vert\Phi\rangle_{xy}$. }
\end{center}
\end{figure}

The basic idea of this symmetric controlled teleportation of an
arbitrary two-particle entangled state is shown in Fig.1. Suppose
that Alice wants to send the state $\vert \Phi\rangle_{xy}$ to one
of the two agents randomly and the receiver can reconstruct the
state only when he/she obtains the help of the other agent, i.e.,
Bob reconstructs it with the control of Charlie's, or vice versa.
To this end, Alice prepares two three-particle GHZ states $\vert
\Psi\rangle_{a_1a_2a_3}$ and $\vert \Psi\rangle_{b_1b_2b_3}$,
\begin{eqnarray}
\vert \Psi\rangle_{a_1a_2a_3}=\vert
\Psi\rangle_{b_1b_2b_3}=\frac{1}{\sqrt{2}}(\vert 000\rangle+\vert
111\rangle),
\end{eqnarray}
and she sends the particles $a_2$ and $b_2$ to Charlie, and $a_3$
and $b_3$ to Bob. The state of the composite quantum system
composed of the eight particles $x$, $y$, $a_1$, $a_2$, $a_3$,
$b_1$, $b_2$,  and $b_3$ can be written as
\begin{eqnarray}
\vert \Psi\rangle_{s}&\equiv& \vert \Phi\rangle_{xy} \otimes \vert
\Psi\rangle_{a_1a_2a_3} \otimes \vert \Psi\rangle_{b_1b_2b_3}\nonumber\\
&=&(a\vert 00\rangle + b \vert 01\rangle + c\vert 10\rangle +
d\vert 11\rangle)_{xy} \nonumber\\
&\otimes& \frac{1}{\sqrt{2}}(\vert 000\rangle+\vert
111\rangle)_{a_1a_2a_3} \nonumber\\
&\otimes& \frac{1}{\sqrt{2}}(\vert 000\rangle+\vert
111\rangle)_{b_1b_2b_3}.
\end{eqnarray}
Alice performs Bell-basis measurements on the particles $x$ and
$a_1$, and $y$ and $b_1$, respectively, and then publishes the
outcomes. If Bob wants to reconstruct the state $\vert
\Phi\rangle_{xy}$, Charlie does the Bell-basis measurement on her
particles $a_2$ and $b_2$, or vice versa. Without loss of
generalization, we assume that Bob will obtain the original state
with the help of Charlie, shown in Fig.1.

In fact, Bob can only get an EPR-class
entangled state, i.e., $\vert
\Phi\rangle_{u}=\alpha \vert uv\rangle + \beta
\vert \bar{u}\bar{v}\rangle$ ($u,v\in \{0, 1\}$ and $\vert
\alpha\vert^2+\vert \beta\vert^2=1$), similar to those in Refs.
\cite{TT1,TT2,dengqss} if Alice and Charlie perform Bell-basis
measurements on the composite quantum system $\vert
\Psi\rangle_{s}$ directly. For example, if the results of the
Bell-basis measurements are $\vert \phi^+\rangle_{xa_1}$, $\vert
\phi^+\rangle_{yb_1}$ and $\vert \phi^+\rangle_{a_2b_2}$, then the
 particles $a_3$ and $b_3$ are in the state
\begin{eqnarray}
\vert \Psi\rangle_{a_3b_3}&=& _{a_2b_2}\langle \phi^+\vert \otimes
_{yb_1}\langle \phi^+ \vert \otimes_{xa_1}\langle \phi^+\vert
\Phi\rangle_{s}\nonumber\\
&=&\frac{1}{4\sqrt{2}}(a\vert 00\rangle + d\vert
11\rangle)\Rightarrow (\alpha\vert 00\rangle + \beta \vert
11\rangle).
\end{eqnarray}
It is just a superposition of the two product states $\vert
00\rangle$ and $\vert 11\rangle$. Fortunately, the case will be
changed with just a little of modification. Instead of sending the
three particles in the state $\vert
\Psi\rangle_{b_1b_2b_3}=\frac{1}{\sqrt{2}}(\vert 000\rangle +
\vert 111\rangle)$ directly, Alice transfers it into $\vert
\Psi'\rangle_{b_1b_2b_3}=\frac{1}{\sqrt{2}}(\vert +x+x+x\rangle +
\vert -x-x-x\rangle)$ with a H operation on each particle. Then
the joint state of the composite quantum system is transferred to
be
\begin{eqnarray}
\vert \Psi \rangle_{joint} &\equiv& \vert \Phi\rangle_{xy}\otimes
\vert \Psi\rangle_{a_1a_2a_3} \otimes \vert \Psi'\rangle_{b_1b_2b_3}\nonumber\\
&=&\frac{1}{2}(a\vert 00\rangle + b \vert 01\rangle + c\vert
10\rangle + d\vert 11\rangle)_{xy}\nonumber\\
&\otimes& (\vert 000\rangle + \vert 111\rangle)_{a_1a_2a_3}\\
&\otimes& (\vert +x+x+x\rangle + \vert
-x-x-x\rangle)_{b_1b_2b_3}\nonumber. \label{Jstate}
\end{eqnarray}
Using the decomposition into Bell states, we can get the relation
between the measurement results (i.e., $R_{xa_1}$, $R_{yb_1}$,
$R_{a_2b_2}$) and the final state of the two particles $a_3$ and
$b_3$, $\vert \Phi\rangle_{a_3b_3}$, shown in Table I.

\begin{widetext}
\begin{center}
\begin{table}[!h]
\label{table1} \caption{The relation between the unitary
operations and the results $R_{xa_1}$, $R_{yb_1}$, and
$R_{a_2b_2}$ in the case that each of Alice, Bob, and Charlie
keeps one of the three particles in each GHZ state.
$\Phi_{a_3b_3}$ is the state of the two particles held by Bob
after all the Bell-basis measurements are done by the sender Alice
and the controller Charlie.}
\begin{tabular}{ccccccc|cccccc}\hline
$V_{xa_1}$  & & $V_{total}$& & $P_{yb_1}$ &  & $P_{total}$ & & & &
$\Phi_{a_3b_3}$ & & operations\\\hline
 0  & & 0 & & $+$ & & $+$ & &  & & $a\vert
00\rangle + b\vert 01\rangle + d\vert 10\rangle + c \vert
11\rangle$ & & $U_0\otimes U_0 + CNot$ \\
 0  & & 0 & & $+$ & & $-$ & &  & &
$a\vert 00\rangle + b\vert 01\rangle -d \vert 10\rangle - c \vert
11\rangle$ & & $U_1\otimes U_0 + CNot$
\\
0  & & 0 & & $-$ & & $+$ & &  & & $a\vert 00\rangle - b\vert
01\rangle + d\vert 10\rangle - c \vert 11\rangle$ & & $U_0\otimes
U_1 + CNot$
\\
 0  & & 0 & & $-$ & & $-$ & &  & & $a\vert
00\rangle - b\vert 01\rangle - d\vert 10\rangle + c \vert
11\rangle$ & & $U_1\otimes U_1 + CNot$
\\
 0  & & 1 & & $+$ & & $+$ & &  & & $b\vert
00\rangle + a\vert 01\rangle + c  \vert 10\rangle +d\vert
11\rangle$ & & $U_0\otimes U_2 + CNot$
\\
 0  & & 1 & & $+$ & & $-$ & &  & & $b\vert
00\rangle + a\vert 01\rangle - c\vert 10\rangle - d \vert
11\rangle$ & & $U_1\otimes U_2 + CNot$
\\
 0  & & 1 & & $-$ & & $+$ & &  & & $b\vert
00\rangle - a\vert 01\rangle + c\vert 10\rangle - d \vert
11\rangle$ & & $U_0\otimes U_3 + CNot$
\\
 0  & & 1 & & $-$ & & $-$ & &  & & $b\vert
00\rangle - a\vert 01\rangle - c\vert 10\rangle + d \vert
11\rangle$ & & $U_1\otimes U_3 + CNot$
\\
 1  & & 0 & & $+$ & & $+$ & &  & & $d\vert
00\rangle + c\vert 01\rangle + a\vert 10\rangle + b  \vert
11\rangle$ & & $U_2\otimes U_0 + CNot$
\\
 1  & & 0 & & $+$ & & $-$ & &  & & $d\vert
00\rangle + c\vert 01\rangle - a \vert 10\rangle - b \vert
11\rangle$ & & $U_3\otimes U_0 + CNot$
\\
 1  & & 0 & & $-$ & & $+$ & &  & & $d\vert
00\rangle - c\vert 01\rangle + a\vert 10\rangle - b \vert
11\rangle$ & & $U_2\otimes U_1 + CNot$
\\
 1  & & 0 & & $-$ & & $-$ & &  & & $d\vert
00\rangle - c\vert 01\rangle - a\vert 10\rangle + b \vert
11\rangle$ & & $U_3\otimes U_1 + CNot$
\\
 1  & & 1 & & $+$ & & $+$ & &  & & $c\vert
00\rangle + d\vert 01\rangle + b\vert 10\rangle + a \vert
11\rangle$ & & $U_2\otimes U_2 + CNot$
\\
 1  & & 1 & & $+$ & & $-$ & &  & & $c\vert
00\rangle + d\vert 01\rangle - b\vert 10\rangle - a \vert
11\rangle$ & & $U_3\otimes U_2 + CNot$
\\
 1  & & 1 & & $-$ & & $+$ & &  & & $c\vert
00\rangle - d\vert 01\rangle + b\vert 10\rangle - a \vert
11\rangle$ & & $U_2\otimes U_3 + CNot$
\\
1  & & 1 & & $-$ & & $-$ & &  & & $c\vert 00\rangle - d\vert
01\rangle -b\vert 10\rangle + a \vert 11\rangle$ & & $U_3\otimes
U_3 + CNot$
\\\hline
\end{tabular}
\end{table}
\end{center}
\end{widetext}

\begin{figure}[!h]
\begin{center}
\includegraphics[width=8cm,angle=0]{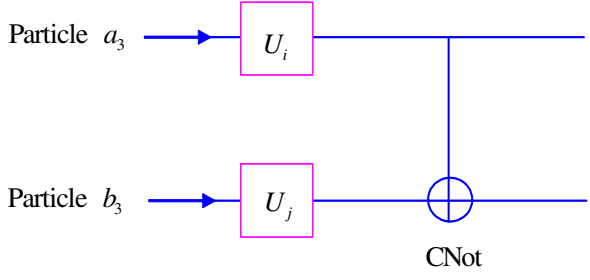} \label{fig2}
\caption{ (Color online) The operations that Bob needs to perform
on the two particles for reconstructing the original entangled
state. $U_i$, $U_j$ $\in$ $\{U_0, U_1, U_2, U_3\}$. }
\end{center}
\end{figure}

Now, let us describe the notations in Table I. Here we define $V$
as the bit value of the Bell state, i.e., $V_{\vert
\phi^\pm\rangle}\equiv 0$, $V_{\vert \psi^\pm\rangle}\equiv 1$.
That is, the bit value $V=0$ if the states of the two particles in
a Bell state are parallel, otherwise $V=1$. $V_{total} \equiv
V_{xa_1}\oplus V_{yb_1} \oplus V_{a_2b_2}$. $P$ denotes the parity
of the result of the Bell-basis measurement on the two-particle
quantum system $R_i \in \{\vert \psi^+\rangle, \vert
\psi^-\rangle, \vert \phi^+\rangle, \vert \phi^-\rangle \}$, i.e.,
$P_{\vert \psi ^\pm\rangle}\equiv\pm$, $P_{\vert \phi
^\pm\rangle}\equiv\pm$ and $P_{total} \equiv \prod_{i=1} \otimes
P_{R_i}=P_{R_{xa_1}}\otimes P_{R_{yb_1}}\otimes P_{R_{a_2b_2}}$;
$\Phi_{a_3b_3}$ is the state of the two particles $a_3$ and $b_3$
after all the Bell-basis measurements are taken by Alice and
Charlie; the unitary operations $U_i\otimes U_j + CNot$ $(i,j \in
\{0,1,2,3\})$ means performing the unitary operation $U_i$ on the
particle $a_3$ and the operation $U_j$ on the particle $b_3$,
respectively, and then taking a controlled-not (CNOT) gate on
those two particles for reconstructing the state $\vert
\Phi\rangle_{xy}$, shown in Fig.2. For example, if the results of
$R_{xa_1}$, $R_{yb_1}$, and $R_{a_2b_2}$ are $\vert
\psi^-\rangle_{xa_1}$, $\vert \phi^-\rangle_{yb_1}$ and $\vert
\psi^-\rangle_{a_2b_2}$, respectively, then $V_{R_{xa_1}}=1$,
$V_{total}=V_{xa_1}\oplus V_{yb_1}\oplus V_{a_2b_2}=1\oplus
0\oplus 1=0$, $P_{yb_1}=-$, $P_{total}=(-)\otimes (-) \otimes
(-)=-$, and Bob first performs the unitary operations $U_3$ and
$U_1$ on the particles $a_3$ and $b_3$,  respectively, and then
does the CNOT operation on those two particles for reconstructing
the state $\vert\Phi\rangle_{xy}$.



Unlike those in Refs. \cite{TT1,TT2,dengqss}, the original
entangled state is an arbitrary one, i.e.,
$\vert\Phi\rangle_{xy}=a\vert 00\rangle + b\vert 01\rangle +
c\vert 10\rangle + d\vert 11\rangle$ is an arbitrary state in the
Hilbert space $\mathcal{H}^2\otimes \mathcal{H}^2$ for two
particles. Another feature in this controlled teleportation is
that the receiver should perform a CNOT gate on the two particles
for recovering the state $\vert\Phi\rangle_{xy}$. Moreover, the
whole quantum source is used to carry useful information and the
efficiency for the qubits $\eta_q\equiv \frac{q_u}{q_t}$
approaches the maximal value $\frac{1}{3}$ as the receiver can
recover the two-qubit entangled state with a six-qubit quantum
source, where $q_u$ is the number of useful qubits and $q_t$ is
the number of qubits used for teleportation.

\section{Controlled teleportation via the control of $n$ agents}

In this section, we will generalize the  method discussed above to
the case that there are $n$ controllers who control the
teleportation of an arbitrary two-particle entangled state
$\vert\Phi\rangle_{xy}=a\vert 00\rangle + b\vert 01\rangle +
c\vert 10\rangle + d\vert 11\rangle$, say Charlie$_i$
$\{i=1,2,..,n\}$, shown in Fig.3.

For the controlled teleportation, Alice prepares two
$(n+2)$-particle GHZ states. The state of the composite quantum
system can be written as
\begin{eqnarray}
\vert \Psi\rangle_{s}&\equiv& \vert \Phi\rangle_{xy} \otimes \vert
\Psi\rangle_{s_1} \otimes \vert \Psi\rangle_{s_2}\nonumber\\
&=&(a\vert 00\rangle + b \vert 01\rangle + c\vert 10\rangle +
d\vert 11\rangle)_{xy} \nonumber\\
&\otimes& \frac{1}{\sqrt{2}}(\prod_{i=1}^{n+2}\vert
0\rangle_{a_i}+\prod_{i=1}^{n+2}\vert 1\rangle_{a_i}) \nonumber\\
&\otimes& \frac{1}{\sqrt{2}}(\prod_{i=1}^{n+2}\vert
+x\rangle_{b_i}+\prod_{i=1}^{n+2}\vert -x\rangle_{b_i}).
\end{eqnarray}
After the Bell-basis measurements on the particles $x$ and $a_1$,
and $y$ and $b_1$, respectively, are done by Alice, the state of
the subsystem (without being normalized) becomes
\begin{eqnarray}
\Psi_{sub}&=& (\prod_{i=2}^{n+2}\vert
0\rangle_{a_i})\otimes[\alpha (\prod_{i=2}^{n+2}\vert
+x\rangle_{b_i})\nonumber\\
&+&\beta(\prod_{i=2}^{n+2}\vert
-x\rangle_{b_i})]+(\prod_{i=2}^{n+2}\vert
1\rangle_{a_i})\nonumber\\
&\otimes& [\gamma (\prod_{i=2}^{n+2}\vert
+x\rangle_{b_i})+\delta(\prod_{i=2}^{n+2}\vert -x\rangle_{b_i})].
\end{eqnarray}
The relation between the numbers $\alpha$, $\beta$, $\gamma$,
$\delta$ and the results $R_{xa_1}$, $R_{yb_1}$ is shown in Table
II.

\begin{figure}[!h]
\begin{center}
\includegraphics[width=8cm,angle=0]{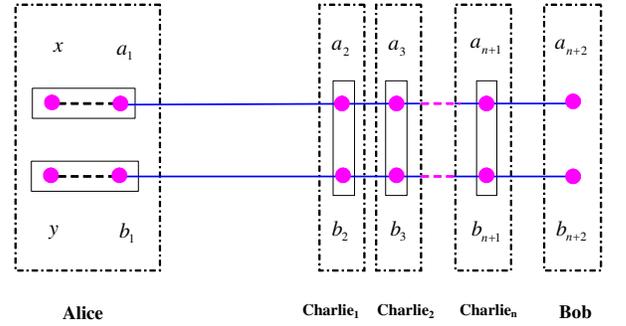} \label{fig3}
\caption{ (Color online) The principle of the controlled
teleportation of an arbitrary two-particle entangled state in the
case that there are $n$ controllers. The rectangles represent the
Bell-basis measurements done by Alice or the controllers;
Charlie$_i$ are the $n$ controllers in the $n+1$ agents; Bob is
just the agent who will obtain the original entangled state with
unitary operations. }
\end{center}
\end{figure}

We can use a common formula to represent the Bell-basis
measurements done by the $n$ controllers, Charlie$_s$, i.e.,
\begin{eqnarray}
M\equiv (\langle\psi^-\vert)^{n-m-l-k}\otimes
(\langle\psi^+\vert)^{m}\otimes (\langle\phi^-\vert)^{l}\otimes
(\langle\phi^+\vert)^{k}.
\end{eqnarray}
It means that the numbers of the controllers who obtain the
results of Bell-basis measurements $\vert \phi^+\rangle$, $\vert
\phi^-\rangle$, $\vert \psi^+\rangle$ and $\vert \psi^-\rangle$
are $k$, $l$, $m$, and $n-k-l-m$, respectively. Because of the
symmetry,  who obtains the result $\vert \phi^+\rangle$ is not
important for the final state $\vert \Psi\rangle_{a_{n+2}b_{n+2}}$
of the two particles $a_{n+2}$ and $b_{n+2}$, but the number. The
operation $M$ represents all the possible results of the
Bell-basis measurements of the $n$ controllers with the parameters
$k$, $l$, and $m$. The final state $\vert
\Psi\rangle_{a_{n+2}b_{n+2}}$ can be obtained by means of
performing the operation $M$ on the state of the subsystem
$\Psi_{sub}$:
\begin{widetext}
\begin{eqnarray}
\Psi_{a_{n+2}b_{n+2}}&=& M\{(\prod_{i=2}^{n+2}\vert
0\rangle_{a_i})\otimes[\alpha (\prod_{i=2}^{n+2}\vert
+x\rangle_{b_i})+\beta(\prod_{i=2}^{n+2}\vert
-x\rangle_{b_i})]+(\prod_{i=2}^{n+2}\vert
1\rangle_{a_i})\otimes[\gamma (\prod_{i=2}^{n+2}\vert
+x\rangle_{b_i})+\delta(\prod_{i=2}^{n+2}\vert
-x\rangle_{b_i})]\}\nonumber\\
&=&\frac{1}{2^{n}}\{\vert 0\rangle_{a_{n+2}}[\alpha \vert
+x\rangle+(-1)^{n-l-k}\beta\vert
-x\rangle]_{b_{n+2}}+(-1)^{n-m-k}\vert
1\rangle_{a_{n+2}}[\gamma\vert +x\rangle+(-1)^{k+l}\delta\vert
-x\rangle]_{b_{n+2}}\}\nonumber\\
&=&\frac{\sqrt2}{2^{n+1}}\{[\alpha+(-1)^{n-l-k}\beta]\vert
00\rangle + [\alpha-(-1)^{n-l-k}\beta]\vert 01\rangle\nonumber\\
&+& (-1)^{n-m-k}[\gamma+(-1)^{k+l}\delta]\vert 10\rangle +
(-1)^{n-m-k}[\gamma-(-1)^{k+l}\delta]\vert 11\rangle\}.
\end{eqnarray}
\end{widetext}
This is just the final state $\Psi_{f}$ without being normalized:
\begin{eqnarray}
\Psi_{f}&=&[\alpha+(-1)^{n-l-k}\beta]\vert 00\rangle +
[\alpha-(-1)^{n-l-k}\beta]\vert 01\rangle \nonumber\\
&+& (-1)^{n-m-k}[\gamma+(-1)^{k+l}\delta]\vert 10\rangle \nonumber\\
&+& (-1)^{n-m-k}[\gamma-(-1)^{k+l}\delta]\vert 11\rangle.
\end{eqnarray}

Similar to that in the case with one controller, let us define
\begin{eqnarray}
V_{total}\equiv \sum_{i} \oplus V_{R_i}, P_{total} \equiv
\prod_{i} \otimes P_{R_i},
\end{eqnarray}
where $V_{R_i}$, $P_{R_i}$ are the bit values and the parities of
the results of the Bell-basis measurements done by Alice or the
controllers, respectively, see them in section II.

\begin{widetext}
\begin{center}
\begin{table}[!h]
\label{table2} \caption{The relation between the values of
$\alpha$, $\beta$, $\gamma$, $\delta$ and the results of the
Bell-basis measurements on the particles $x$ and $a_1$, $y$ and
$b_1$. }
\begin{tabular}{cccccccc|ccccccccc}\hline
$V_{xa_1}$  & & $V_{yb_1}$ & & $P_{xa_1}$& & $P_{yb_1}$ & & & &
$\alpha$ &  & $\beta$ & & $\gamma$& & $\delta$
\\\hline
0  & & 0& & $+$ & & $+$ & & & & $+(a+b)$ & & $+(a-b)$  & &
$+(c+d)$ & & $+(c-d)$
\\
0  & & 0& & $+$ & & $-$  & & & & $+(a-b)$ & & $+(a+b)$  & &
$+(c-d)$ & & $+(c+d)$
\\
0  & & 0& & $-$ & & $+$ & & & & $+(a+b)$ & & $+(a-b)$  & &
$-(c+d)$ & & $-(c-d)$
\\
0  & & 0& & $-$ & & $-$ & & & & $+(a-b)$ & & $+(a+b)$  & &
$-(c-d)$ & & $-(c+d)$
\\
0  & & 1& & $+$ & & $+$  & & & & $+(a+b)$ & & $-(a-b)$ & &
$+(c+d)$ & & $-(c-d)$
\\
0  & & 1& & $+$ & & $-$ & & & & $+(a-b)$ & & $-(a+b)$ & & $+(c-d)$
& & $-(c+d)$
\\
0  & & 1& & $-$ & & $+$ & & & & $+(a+b)$ & & $-(a-b)$ & & $-(c+d)$
& & $+(c-d)$
\\
0  & & 1& & $-$ & & $-$ & & & & $+(a-b)$ & & $-(a+b)$ & & $-(c-d)$
& & $+(c+d)$
\\
1  & & 0& & $+$ & & $+$ & & & & $+(c+d)$ & & $+(c-d)$  & &
$+(a+b)$ & & $+(a-b)$
\\
1  & & 0& & $+$ & & $-$ & & & & $+(c-d)$ & & $+(c+d)$  & &
$+(a-b)$ & & $+(a+b)$
\\
1  & & 0& & $-$ & & $+$ & & & & $-(c+d)$ & & $-(c-d)$  & &
$+(a+b)$ & & $+(a-b)$
\\
1  & & 0& & $-$ & & $-$ & & & & $-(c-d)$ & & $-(c+d)$  & &
$+(a-b)$ & & $+(a+b)$
\\
1  & & 1& & $+$ & & $+$ & & & & $+(c+d)$ & & $-(c-d)$  & &
$+(a+b)$ & & $-(a-b)$
\\
1  & & 1& & $+$ & & $-$ & & & & $+(c-d)$ & & $-(c+d)$  & &
$+(a-b)$ & & $-(a+b)$
\\
1  & & 1& & $-$ & & $+$ & & & & $-(c+d)$ & & $+(c-d)$  & &
$+(a+b)$ & & $-(a-b)$
\\
1  & & 1& & $-$ & & $-$ & & & & $-(c-d)$ & & $+(c+d)$  & &
$+(a-b)$ & & $-(a+b)$
\\\hline
\end{tabular}
\end{table}

\begin{table}[!h]
\label{table3} \caption{The relation between the results of the
Bell-basis measurements
 and the state $\Psi_{f}$ when the number of the controllers is even. }
\begin{tabular}{ccccccc|cccccc}\hline
$V_{xa_1}$  & & $V_{total}$ & & $P_{yb_2}$ &  & $P_{total}$ & & &
& $\Psi_{f}$& & operations
\\\hline
 0  & & 0 & & $+$ & & $+$ & &  & & $a\vert
00\rangle + b\vert 01\rangle + c\vert 10\rangle + d \vert
11\rangle$ & & $U_0 \otimes U_0$\\
 0  & & 0 & & $+$ & & $-$ & &  & &
$a\vert 00\rangle + b\vert 01\rangle - c\vert 10\rangle - d \vert
11\rangle$ & & $U_1 \otimes U_0$
\\
0  & & 0 & & $-$ & & $+$ & &  & & $a\vert 00\rangle - b\vert
01\rangle - c\vert 10\rangle + d \vert 11\rangle$ & & $U_1 \otimes
U_1$
\\
 0  & & 0 & & $-$ & & $-$ & &  & & $a\vert
00\rangle - b\vert 01\rangle + c\vert 10\rangle - d \vert
11\rangle$ & & $U_0 \otimes U_1$
\\
 0  & & 1 & & $+$ & & $+$ & &  & & $b\vert
00\rangle + a\vert 01\rangle + d\vert 10\rangle + c \vert
11\rangle$ & & $U_0 \otimes U_2$
\\
 0  & & 1 & & $+$ & & $-$ & &  & & $b\vert
00\rangle + a\vert 01\rangle - d\vert 10\rangle - c \vert
11\rangle$ & & $U_1 \otimes U_2$
\\
 0  & & 1 & & $-$ & & $+$ & &  & & $b\vert
00\rangle - a\vert 01\rangle - d\vert 10\rangle + c \vert
11\rangle$ & & $U_1 \otimes U_3$
\\
 0  & & 1 & & $-$ & & $-$ & &  & & $b\vert
00\rangle - a\vert 01\rangle + d\vert 10\rangle - c \vert
11\rangle$ & & $U_0 \otimes U_3$
\\
 1  & & 0 & & $+$ & & $+$ & &  & & $d\vert
00\rangle + c\vert 01\rangle + b\vert 10\rangle + a \vert
11\rangle$ & & $U_2 \otimes U_2$\\
 1  & & 0 & & $+$ & & $-$ & &  & & $d\vert
00\rangle + c\vert 01\rangle - b\vert 10\rangle - a \vert
11\rangle$ & & $U_3 \otimes U_2$\\
 1  & & 0 & & $-$ & & $+$ & &  & & $d\vert
00\rangle - c\vert 01\rangle - b\vert 10\rangle + a \vert
11\rangle$ & & $U_3 \otimes U_3$\\
 1  & & 0 & & $-$ & & $-$ & &  & & $d\vert
00\rangle - c\vert 01\rangle + b\vert 10\rangle - a \vert
11\rangle$ & & $U_2 \otimes U_3$\\
 1  & & 1 & & $+$ & & $+$ & &  & & $c\vert
00\rangle + d\vert 01\rangle + a\vert 10\rangle + b \vert
11\rangle$ & & $U_2 \otimes U_0$\\
 1  & & 1 & & $+$ & & $-$ & &  & & $c\vert
00\rangle + d\vert 01\rangle - a\vert 10\rangle - b \vert
11\rangle$ & & $U_3 \otimes U_0$\\
 1  & & 1 & & $-$ & & $+$ & &  & & $c\vert
00\rangle - d\vert 01\rangle - a\vert 10\rangle + b \vert
11\rangle$ & & $U_3 \otimes U_1$\\
1  & & 1 & & $-$ & & $-$ & &  & & $c\vert 00\rangle - d\vert
01\rangle + a\vert 10\rangle - b \vert 11\rangle$ & & $U_2 \otimes
U_1$\\\hline
\end{tabular}
\end{table}
\end{center}
\end{widetext}

The relation between the state $\Psi_{f}$ and the results
$V_{xa_1}$, $V_{total}$, $P_{yb_2}$, and  $P_{total}$ is shown in
Table III when the number of the controllers $n$ is even. When $n$
is odd, the result is the same as that in Table I with just the
modification of replacing the state $\Phi_{a_3b_3}$ with
$\Psi_{f}$. The results in Table I and III show that the unitary
operations performed on Bob's particles for reconstructing the
state $\vert \Phi\rangle_{xy}$ are different in principle when $n$
is even or odd. In Table III, it is enough for Bob to reconstruct
the state $\vert \Phi\rangle_{xy}$ with the two local unitary
operations, $U_i$ and $U_j$ ($i,j$$\in$$\{0,1,2,3\}$) on the
particles $a_{n+2}$ and $b_{n+2}$, respectively, but he has to do
an additional CNOT operation on the two particles when the number
of the controller is odd, which is different from the other
methods for a controlled teleportation
\cite{CTele1,YanFLteleportation2,CTele2}.


For a secure controlled teleportation of the state $\vert
\Phi\rangle_{xy}$, the controllers need to keep the receiver from
eavesdropping the quantum communication when they set up the
quantum channel, similar to the case in quantum secret sharing.
Surely, the task of the teleportation of an arbitrary two-particle
entangled state can be completed with the combination of the
method for teleporting an arbitrary two-qubit state \cite{Rigolin}
and quantum secure direct communication protocols
\cite{dengqsdc,beige,QOTP,yan,zhangzj,wangcPRA}, similar to the
way that quantum secret sharing for classical information
\cite{DZL} can be finished with quantum-key-distribution protocols
\cite{BB84,Ekert91,BBM92,Gisin,cabello,LongLiu,DLQKD}. This time,
the receiver is only the person who is deterministic in advance,
not an arbitrary man in the $n+1$ agents. Moreover, the total
efficiency $\eta_t$ is not more than that in this symmetric
controlled teleportation protocol, as the classical information
exchanged and the quantum source will increase since the
efficiency of QKD is no more than 1. Here $\eta_t$ is defined as
\cite{cabello,LongLiu}
\begin{equation}
\eta_t=\frac{q_u}{q_t+b_t}, \label{eff1}
\end{equation}
where $b_t$ is the number of classical bits exchanged between the
parties. On the other hand, the multiparticle-entangled states
must be produced in this protocol, which is not easy at present
\cite{Mentanglement1,Mentanglement2,Mentanglement3}. With the
improvement of technology, it may be feasible in the future.

\section{Quantum secret Sharing based on controlled teleportation}

\subsection{Setting up the quantum channel with GHZ states}
It is important for the parties of the communication to set up a
quantum channel with GHZ states securely in both the symmetric
controlled teleporation and quantum secret sharing. The process
for constructing a quantum channel discussed in this paper is
similar to that in Ref. \cite{dengqsdc} for quantum secure direct
communication (QSDC) in which the classical secret is transmitted
directly without creating a private key and then encrypting it.
Another property, as in QSDC
\cite{dengqsdc,beige,QOTP,yan,zhangzj,wangcPRA}, is that the
information about the unknown state $\vert \Phi\rangle_{xy}$
should not be leaked to an unauthorized user, such as a vicious
eavesdropper Eve. It means that the controllers and Eve can get
nothing about the final entangled state even though they eavesdrop
on the quantum communication. If the quantum channel is secure,
no-one can obtain the original state except for the legal receiver
Bob.

If the process for constructing the entangled quantum channel is
secure, then the whole process for communication is secure as
no-one can read out the information about a maximal entangled
quantum system from a part of its \cite{book}. The results in
Secs. II and
 III show that Bob's particles is randomly in one
of the 16 entangled states with the same probability. The
randomness of the outcomes ensures the security of the
communication \cite{book},  as for the classical one-time-pad
cryptosystem \cite{onetime}.

The method for setting up a quantum channel with a sequence of EPR
pairs is discussed in Refs. \cite{guoatom,dengqsdc}.  The approach
can also be used for sharing GHZ states \cite{book}. The way is
just that the legal users determine whether there is an
eavesdropper in the line when they transmit the particles in the
GHZ state and then purify the quantum channel if there is no-one
monitoring the line or the probability for being eavesdropped is
lower than a suitable threshold. The latter can be considered as
quantum privacy amplification with quantum purification
\cite{qpap1,qpap2}. Let us use a three-particle GHZ state as an
example to demonstrate the principle, as in Ref. \cite{dengqss}.
Alice prepares a sequence of GHZ states $\vert GHZ\rangle_{ABC}$.
For each GHZ state, Alice sends the particles $B$ and $C$ to Bob
and Charlie, respectively, and retains the particle $A$:
\begin{eqnarray}
&\vert& GHZ\rangle_{ABC}=\frac{1}{\sqrt{2}}(\vert
000\rangle_{ABC}+\vert 111\rangle_{ABC})\nonumber\\
&=&\frac{1}{2\sqrt{2}}[(\vert+x\rangle_A\vert+x\rangle_B+\vert-x\rangle_A\vert-x\rangle_B)
\vert+x\rangle_C\nonumber\\
&+&(\vert+x\rangle_A\vert-x\rangle_B+\vert-x\rangle_A\vert+x\rangle_B)\vert-x\rangle_C].
\end{eqnarray}
For determining whether there is an eavesdropper in the line when
the particles are transmitted, Alice picks up some of the GHZ
states from the GHZ sequence randomly, and requires Bob and
Charlie to choose the measuring basis $\sigma_z$ or $\sigma_x$ to
measure their particles according to the information published by
Alice. If there is an eavesdropper monitoring the quantum channel,
the error rate of the samples will increase, as in the
Bennett-Brassard-Mermin (BBM92) QKD protocol \cite{BBM92}. If the
error rate is low, Alice, Bob, and Charlie can obtain some private
GHZ states with multiparticle entanglement purification
\cite{qpap1,qpap2}. For preventing a dishonest agent from
eavesdropping with a fake signal, Alice inserts some decoy photons
in the GHZ sequence. The decoy photons can be produced by means
that Alice measures a particle in some three-particle GHZ-state
quantum system with the MB $\sigma_x$ and sends another particle
to a agent and keeps the last one. That is, Alice shares some Bell
states with each agent. If the dishonest agent intercepts these
decoy photons in the GHZ sequence, his action will introduce
inevitably errors in the results of those samples when Alice and
the other agent check eavesdropping with the two measuring bases,
$\sigma_z$ and $\sigma_x$, same as BBM92 QKD protocol
\cite{BBM92}.

\subsection{Quantum secret sharing of a classical secret and
quantum information with controlled teleportation}

Now, let us introduce the way for quantum secret sharing with the
controlled teleportation. There are two main goals in quantum
secret sharing. One is to share classical information, a sequence
of binary numbers, and the other is to share quantum information,
an unknown quantum state. In the former, the quantum state of each
two particles in all the parties of the communication is coded as
a two-bit binary number (the parties store the results of the
measurements on the particles as a classical information). For
instance, they can code the four Bell states $\{
\vert\phi^+\rangle, \vert\phi^-\rangle, \vert\psi^+\rangle,
\vert\psi^-\rangle\}$ as $\{ 0+, 1-, 0-, 1+\}$, respectively. Here
the codes $\{ +, -\}$ can be used to represent the binary numbers
$\{ 0,1 \}$, respectively.  For sharing an unknown quantum state,
the case is similar to that for the controlled teleportation of an
arbitrary two-particle state, and the agents will recover the
unknown state when they collaborate.

For sharing classical information, Alice encodes her message (a
random key or a classical secret) on her two-particle quantum
state. For the convenience of the measurements, it requires that
the final state of Bob's particles can be measured
deterministically if all the controllers, say Charlie$_{i}$
perform the Bell-basis measurements on their particles, as in QSDC
\cite{dengqsdc,beige,QOTP,yan,zhangzj}. In other words, all the
input states should be orthogonal. Then the quantum system
composed of two particles prepared by Alice for coding the
classical information is in one of the four Bell states or in
EPR-class states, i.e.,
\begin{eqnarray}
\vert\Phi'^\pm\rangle_{xy}=\alpha\vert uv\rangle \pm \beta\vert
\bar{u}\bar{v}\rangle,
\end{eqnarray}
where $u$, $v$ $\in$ $\{\vert 0\rangle, \vert 1\rangle, \vert
+x\rangle, \vert -x\rangle\}$. Without loss of generality, we
suppose that the state $\vert \Psi\rangle_c$ prepared for carrying
the classical secret is one of the four Bell states $\{\vert
\phi^\pm\rangle, \vert \psi^\pm\rangle\}$. Alice prepares two GHZ
states $\vert \Psi\rangle_{s_1}$ and $\vert \Psi\rangle_{s_2}$ as
the quantum channel. And the state of the composite quantum system
is
\begin{eqnarray}
\vert \Psi\rangle_{s}&\equiv& \vert \Psi\rangle_c \otimes \vert
\Psi\rangle_{s_1} \otimes \vert \Psi\rangle_{S_2},
\end{eqnarray}
where
\begin{eqnarray}
\vert \Psi\rangle_{s_1}=\frac{1}{\sqrt{2}}(\prod_{i=1}^{n+2}\vert
0\rangle_{a_i}+\prod_{i=1}^{n+2}\vert 1\rangle_{a_i}),
\end{eqnarray}
\begin{eqnarray}\vert
\Psi\rangle_{s_2}=\frac{1}{\sqrt{2}}(\prod_{i=1}^{n+2}\vert
+x\rangle_{b_i}+\prod_{i=1}^{n+2}\vert -x\rangle_{b_i}).
\end{eqnarray}
Alice and Charlie$_i$ all perform Bell-basis measurements on their
particles. Bob will perform Bell-basis measurement on his two
particles $a_{n+2}$ and $b_{n+2}$ when the number of the
controllers, Charlie$_i$, is even, otherwise Bob will take a joint
measurement $\sigma_x \otimes \sigma_z$ on his particles (that is,
he take a $\sigma_x$ measurement on the particle $a_{n+2}$ and
$\sigma_z$ on the particle $b_{n+2}$) as the final states $\Psi_f$
of Bob's particles in these two cases are different.

\begin{table}[!h]
\label{table4}
\begin{center}
\caption{The relation between the results of the Bell-basis
measurements taken by Alice and Charlie$_i$ and the state
$\Psi_{f}$ of the particles $a_{n+2}$ and $\sigma_z$ when the
number of the controllers is even. }
\begin{tabular}{ccc|cc}\hline
 $V_{total}$ &  & $P_{total}$ & &
 $\Psi_{f}$
\\\hline
0 & & $+$ & &  $(U_0 \otimes U_0)\Psi_c$\\
0 & & $-$ & &  $(U_0 \otimes U_1)\Psi_c$\\
1 & & $+$ & &  $(U_0 \otimes U_2)\Psi_c$\\
1 & & $-$ & &  $(U_0 \otimes U_3)\Psi_c$\\
\hline
\end{tabular}
\end{center}
\end{table}

\begin{table}[!h]
\label{table5}
\begin{center}
\caption{The relation between the results of the Bell-basis
measurements
 and the state $\Psi_{f}$ when the number of the controllers is odd. }
\begin{tabular}{ccc|cc}\hline
 $V_{total}$ &  & $P_{total}$ & &
 $\Psi_{f}$
\\\hline
0 & & $+$ & &  $(CNot + U_0 \otimes U_0)\Psi_c$\\
0 & & $-$ & &  $(CNot + U_0 \otimes U_1)\Psi_c$\\
1 & & $+$ & &  $(CNot + U_0 \otimes U_2)\Psi_c$\\
1 & & $-$ & &  $(CNot + U_0 \otimes U_3)\Psi_c$\\
\hline
\end{tabular}
\end{center}
\end{table}

The relation between the final state $\Psi_f$ of Bob's particles
and the original entangled state $\Psi_c$ is shown in Tables IV
and V for the cases that the number of controllers is even and
odd, respectively. When the $n$ controllers and Bob want to
reconstruct the classical secret, they collaborate to decode the
message with the information published by Alice, according to
Tables IV and  V.

The difference between this QSS protocol for a classical secret
and the symmetric multiparty- controlled teleportation discussed
above is that the input states are orthogonal and all the agents
take the measurements on their particles in the former.

A piece of quantum information can be an arbitrary state of a
quantum system. For a two-particle quantum state, an arbitrary
state can be an entangled state in the general formal shown in Eq.
(\ref{unknownstate}). With the symmetric multiparty-controlled
teleportation discussed above, quantum secret sharing for an
entangled state is easily implemented in principle in the same
way. Moreover, each of the $n+1$ agents can act as the person who
will reconstruct the quantum information with the help of the
others in the network.

\section{Discussion and summary}
In the symmetric multiparty-controlled teleportation, the second
GHZ state is prepared along the $x$ direction. It can also be
prepared along the $z$ direction,like the first GHZ state, i.e.,
\begin{eqnarray}
\vert \Psi\rangle_{s}&\equiv& \vert \Phi\rangle_{xy} \otimes \vert
\Psi\rangle_{s_1} \otimes \vert \Psi\rangle_{s_2}\nonumber\\
&=&\frac{1}{\sqrt{2}}(a\vert 00\rangle + b \vert 01\rangle +
c\vert 10\rangle + d\vert 11\rangle)_{xy} \nonumber\\
&\otimes& \frac{1}{\sqrt{2}}(\prod_{i=1}^{n+2}\vert
0\rangle_{a_i}+\prod_{i=1}^{n+2}\vert 1\rangle_{a_i}) \nonumber\\
&\otimes& \frac{1}{\sqrt{2}}(\prod_{i=1}^{n+2}\vert
0\rangle_{b_i}+\prod_{i=1}^{n+2}\vert 1\rangle_{b_i}).
\end{eqnarray}
At this time, Alice and $n-1$ controllers do the Bell-basis
measurements directly on their particles, and the last controller,
say Charlie$_{n}$ first takes a H operation on  her second
particle $b_{n+1}$ and then performs the Bell-basis measurement on
her two particles. Similar to the case above, Bob can also recover
the original entangled state with the unitary operations. As for
sharing of a classical secret, Alice can also prepare the two GHZ
states along the $z$ direction, and all the persons in the
communication perform Bell-basis measurements on their particles
without the $H$ operation.

In summary, we present a method for symmetric
multiparty-controlled teleportation of an arbitrary two-particle
entangled state with two GHZ states and Bell-basis measurements.
Any one in the $n+1$ agents can reconstruct the original entangled
state with the help of the other $n$ agents in the network and the
information published by the sender, Alice. To this end Alice
prepares two $(n+2)$-particle GHZ states along the $z$ direction
and the $x$ direction, respectively. When the number of the
controllers is even, the receiver, say Bob, need only perform two
local unitary operations on his particles to obtain the original
entangled state with the help of the $n$ controllers in the
network; otherwise, he has to do a CNOT operation on his particles
in addition to the local unitary operations. This method for a
symmetric multiparty-controlled teleportation can also be used to
share classical information and an arbitrary two-particle state
with just a little modification. As the whole quantum source is
used to carry the useful quantum information, the efficiency for
qubits approaches the maximal value and the procedure for
controlled teleportation is an optimal one.

\section*{ACKNOWLEDGMENTS}

F.G.D is very grateful to Professor Gui Lu Long for his help. This
work is supported by the National Natural Science Foundation of
China under Grants No. 10447106, No. 10435020, No. 10254002 and
No. A0325401.

\end{document}